\documentclass[12pt]{article}
\usepackage{a4}
\usepackage{epsfig,url}
\usepackage{lineno}

\begin{document}
\title{Swedish Beams\\ The Story of Particle Accelerators in Sweden}
\author{Volker Ziemann, Uppsala University, Uppsala, Sweden}
\date{\today}
\maketitle
\begin{abstract}\noindent
  Even though many of the experiments leading to the standard model of particle physics
  were done at large accelerator laboratories in the US and at CERN\footnote{For a brief history
    of this story see: V. Ziemann, {\em Beams---the Story of Particle Accelerators and the
      Science they Discover,} Copernicus books, Springer 2024.}, many exciting developments
  happened in smaller national facilities all over the world. In this report we highlight
  the history of accelerator facilities in Sweden which was home to the highest-energy
  cyclotron in Europe for some time in the early 1950s.
\end{abstract}
%
%
After the concept of a linear accelerator was born in Stockholm\footnote{Gustaf Ising,
  {\em Prinzip Einer Methode Zur Herstellung Von Kanalstrahlen Hoher Voltzahl,}
  Arkiv f\"or matematik, astronomi och fysik 18 (1924) 1 (in German).}, many of the
subsequent developments happend in the US, where Lawrence invented the cyclotron in 1930
and the first synchrotrons appeared after World War II. But these developments did not
go unnoticed in Sweden. Already 1937, two years before Lawrence was awarded his Nobel
prize, construction of the first cyclotron in Sweden got under way
in Stockholm\footnote{For a brief overview over Swedish accelerators, see: S. Kullander,
  {\em Swedish accelerators take a look at the past,} CERN Courier, April 2000, available online
from \url{https://cerncourier.com/a/swedish-accelerators-take-a-look-at-the-past/}.}.
\subsection*{Stockholm}
Soon after Manne Siegbahn\footnote{Manne Siegbahn received the Nobel Prize for Physics
  in 1924 "for his discoveries and research in the field of X-ray spectroscopy." More
  details are available online from \url{https://www.nobelprize.org/prizes/physics/1924/summary/}.}
became director for the Nobel Institute in Stockholm he acquired funds to build the
first Swedish accelerator, a cyclotron with a diameter of 80\,cm\footnote{For a
  historical, though somewhat technical account, see: H. Atterling, {\em Cyklotronens
    verkningss\"att och anv\"andning,} Teknisk Tidskrift, \AA rg\aa ng 76, 1946 p. 733,
  available online from \url{https://runeberg.org/tektid/1946/0745.html} (in Swedish).}.
By 1941 this machine delivered deuterons with energies up to 7\,MeV that were used to
irradiate various elements to make them radioactive. These activated materials then
emit radiation used to explore various substances, both for fundamental science, but
also to a large extent for biological and medical purposes. This accelerator operated
until 1977.
\par
A few years after the cyclotron started to deliver beams, plans to construct a larger
one emerged such that from 1947 onwards a second cyclotron with a diameter of
225\,cm was built\footnote{A detailed desciption of the 225\,cm cyclotron can be found
  in: H. Atterling, {\em The acceleration of heavy ions in the Stockholm 225\,cm cyclotron,}
  Arkiv f\"or Fysik 15 (1959) 531.}. Like in all accelerators at the time, the electric power
to excite the massive magnets was produced by a motor-driven generator. The communal power
grid was not yet up to the task. In 1951 this cyclotron was taken into operation and
delivered 25\,MeV deuterons and several other ions, such as carbon, to experimenters.
They used the beams to excite the nuclei of many elements. Subsequently observing the
emitted secondary radiation gave them a way to understand the structure of the excited
nuclei. The accelerator was running into the mid-1980s when the space was needed to
construct a synchrotron, called CRYRING.
\par
Part of the new facility's name was derived from one of the particle sources that
were available to produce many different beams; the ``CRYogenic Stockholm Ion Source''
CRYSIS provided highly ionized particles with most of the electrons ripped off.
These particles were accelerated in a so-called radio-frequency quadrupole and
injected into the synchrotron that was dubbed CRYRING\footnote{For an overview see:
  K. Abrahamsson et al., {\em The First Year with Electron Cooling at CRYRING,}
  Proceedings of the 1993 Particle Accelerator Conference in Washington,  p. 1735;
  available online from \url{https://accelconf.web.cern.ch/p93/PDF/PAC1993_1735.PDF}.}.
It had a circumference of 52\,m and was equipped with an electron cooler that was
used to improved the quality of the stored beams. But it also served as a source
of electrons from which the stored beams could replace the ripped-off electrons.
Apart from ionized atoms, CRYRING also
stored ionized molecules that are composed of several atoms. Picking up an electron
in the cooler often makes these molecules unstable such that they break up in a
process called ``dissociative recombination.'' These processes are highly relevant
to understand the creation of molecules in outer space and to understand the
mechanisms in industrial processes, for example, when etching semiconductors
with fluorocarbons. Many other experiments in atomic, molecular and nuclear
physics, as well as contributions to increase the understanding of beams in
accelerators were done until 2012. At this time the CRYRING was packed into boxes and
moved to the FAIR facility in Germany\footnote{The FAIR facility is described in
  a fortcoming report on ``German Beams.''}, where it enjoys a second life as a
``Facility for Low-energy Antiproton and Ion Research,'' abbreviated FLAIR.
\par
In parallel with operating CRYRING ideas emerged to build an electrostatic storage
ring consisting of two separate rings which was named DESIREE\footnote{DESIREE is
  an acronym derived from ``Double ElectroStatic Ion Ring ExpEriment.'' Information
  about the facility is available from \url{https://www.desiree-infrastructure.com/}.}
that was ready for beams in 2015. Instead of using magnetic fields to guide and focus
the beams, DESIREE uses electric fields, which is advantageous for storing very heavy
beams at very low energies. Even very large molecules, such as Buckminsterfullerenes
made of sixty carbon atoms, can be stored this way. Storing delicate molecules is
made possible by enclosing both rings side-by-side in a $5\,$m$\times 3\,$m$\times 1\,$m
evacuated box that is cooled to extremely low temperatures around 13\,K above absolute
zero\footnote{The absolute zero temperature on the Celsius scale is -273\,$^o$C.}.
The latter prevents heat radiation from disturbing the beams. Inside the box a positively
charged beam travels in one ring and a negatively charged beam in the other ring.
Along a common straight line these beams overlap and move in the same direction,
which allows them to exchange electrons just as
the electron cooler was used to donate electrons to positively charged beams. Only
in DESIREE two rather heavy beams are involved. Remarkably, these experiments tell
the experimeters about the evolution of stars and the light they emit. Apart from
merging beams to exchange electrons, they can also be excited by lasers shining their
light onto the beams. This probes the electronic shell structure of the atoms or
molecules that make up the beams.
\par
Already from 1945 onwards, Olle Wernholm and colleagues constructed electron accelerators
at the Royal Institute of Technology (abbreviated KTH for ``Kungliga Tekniska H\"ogskolan''),
another university in Stockholm. The first machine was a betatron with a diameter of 13\,cm
that accelerated electron beams to 2\,MeV. Soon a sequence of larger electron accelerators
followed, both synchrotrons and microtrons. In the latter the beams are recirculated
with magnets to repeatedly gain energy in a short linear accelerator. An early synchrotron
actually became the nucleus of what later became Maxlab in Lund. This ``Ur-MAX'' was
a 35\,MeV electron synchrotron that was handed over to Lund University in 1953. But more
on that later. 
\par
First, let's see how accelerators evolved in Uppsala.
\subsection*{Uppsala}
Already in the 1940s one of The(odor) Svedberg's\footnote{The Svedberg received the
  Nobel Prize in Chemistry in 1926 ``for his work on disperse systems'' that was based
  on his contributions to develop an ultra-centrifuge. More details are available
  online from \url{https://www.nobelprize.org/prizes/chemistry/1926/summary/}.} students
had built a small accelerator, a
neutron generator that induced radioactivity in a large variety of other
substances\footnote{A more detailed description of the activities in Uppsala
  from planning the cyclotron to the turn of the millennium is given in:
  Torsten Lindqvist, {\em Gustaf Werners institut 50 \aa r}, Acta Universitatis
Upsaliensis, Uppsala, 1999 (in Swedish).}. One of the main customers was John Naeslund,
a professor of gynecology in Uppsala who used them as tracers to study the mobility
of fluids in the human body. He soon requested special substances that the small
machine was unable to produce. Thus, by 1945 the idea was hatched to build a
larger accelerator, a cyclotron. Luckily Naeslund's wife had connections with
Gustaf Werner, the owner of a large clothing factory in Gothenburg and one of
Sweden's wealthiest men.  Moreover, Werner was known to be a philanthropist and
generous sponsor of worthy causes. Very quickly he realized the great potential
of Svedberg and Naeslund's proposal and agreed to finance the new cyclotron.
His main objective was probably to boost nuclear physics and biomedical applications,
but he also hoped that it might help to analyze the nuclear chemistry of synthetic
fibers used for clothing; nylon stockings were all the rage at the time. In order
to optimize taxation, Werner ``ordered'' the analysis of textiles from Svedberg
and would pay for the accelerator to do it.
\par
At the same time, one of Svedberg's colleagues was in the US. Right
after the financing was secured, he was instructed to visit labs operating
cyclotrons and ``collect information about recent installations.'' At Lawrence's lab in
Berkeley he learned about the recently discovered principle of phase
focusing\footnote{Phase focusing is one of the great ideas that provided a quantum leap
  for the performance of accelerators. It is highlighted in {\em Beams} mentioned in
  the first footnote.} that would make much higher energies possible than could be
reached with conventional cyclotrons. Moreover, he could pick up many drawings of
components and soon the decision was reached to build a state-of-the-art synchrocyclotron
in Uppsala. Before long construction started and by 1951 Uppsala had the most powerful
cyclotron in Western Europe accelerating protons to energies close to 200\,MeV.
\par
Irradiating synthetic materials was actually done as one of the earliest experiments.
One of the results showed that exposing plastic to accelerated protons increased the
melting temperature. After Svedberg delivered the final report, ownership of the cyclotron
was handed over to Uppsala University. In the following years the beams were used to make
samples of chemical elements radioactive by inserting them into the cyclotron. Many
of those samples were used for biomedical applications, just as Naeslund had asked for.
Others were irradiated to study the nuclear chemistry of the samples by analyzing their
decay products. A few years later protons could be extracted from the cyclotron and
directed to experimental areas. There the experimenters observed how an accelerated proton
recoils from a nucleus and knocks out other protons. In this way much was learned
about the forces acting inside nuclei. Likewise, crashing the proton beams into a
lithium target produces neutrons. They were later used to understand the influence of cosmic
radiation on electronic components on-board aircraft and spaceships. Already from the
mid-1950s the beam's effect on biological tissue and on tumors was explored and
in 1957 the first patient was treated for cancer in the uterus. Remarkably, over the
sixty years until the cyclotron's retirement in 2015 more than 1500 patients were
treated. At that time construction of the new Skandion clinic adjacent to the university
hospital was complete and cancer treatment continued with a modern cyclotron that
delivers protons with energies up to 230\,MeV.
\par
In the early 1980s, and after having served in two previous experiments, a set of forty
10-ton magnets, was no longer needed at CERN in Geneva. Soon ideas emerged to use
them for a storage ring that was later dubbed CELSIUS\footnote{CELSIUS is an acronym
  for ``Cooling with ELectrons and Storing of Ions from the Uppsala Synchrocyclotron.''}.
The venerable cyclotron served as an injector and by 1989 first proton beams were
circulating with energies above 1300\,MeV in the 82\,m long ring. A little later an
electron cooler, constructed at KTH in Stockholm, was installed in CELSIUS. It
dramatically improved the beam quality for two experimental detectors in CELSIUS.
The WASA\footnote{WASA is an acronym derived from ``Wide Angle Shower Apparatus.''}
detector was built around a so-called pellet target generator that shot tiny spheres
of frozen hydrogen through the circulating beam. WASA then recorded flashes caused by
the collision products in more than one thousand crystal detectors that surrounded the
collision point. These experiments were important to improve our understanding of
the nuclear forces. But CELSIUS could also store heavy ion beams, for example, made
of nitrogen or argon ions. These beams were mostly used by the second experiment
called CHIC\footnote{CHIC is an acronym derived from ``CELSIUS Heavy Ion Collaboration.''}.
The CHIC collaboration built their detector around a target station that streamed a gas
jet made of heavy atoms through the circulating beam. Analyzing the escaping collision
products then provided information about the behavior of more complex nuclei. CELSIUS
operated until 2005 at which point the accelerator was disassembled. WASA, however,
found a new home in the storage ring COSY in Germany.
\par
In order to strengthen the capabilities to perform experiments with lower-energy
particles, a tandem accelerator was purchased in the late 1960s. It was based on a
6\,MV van de Graaff accelerator and operated from 1970 until the physics department
moved to the new \AA ngstr\"om laboratory in the late 1990s, when a more modern
tandem accelerator was purchased. Today it is mostly used for radio-carbon dating
and for analyzing materials by observing particles or X-rays that are emitted from
samples as a consequence of the impinging beams. Later the park of accelerators
grew by adding a dedicated system for carbon dating and an ion implanter. The latter
is used to modify the materials used in semiconductors, solar cells, and batteries.
\par
Already in the 1960s the presence of the cyclotron triggered the formation of
a spin-off company. Scanditronix delivered accelerators, both cyclotrons, microtrons,
and synchrotrons as well as accelerator components such as magnets and equipment for
dosimetry to customers worldwide\footnote{For an overview, see: B. Anderberg,
  {\em Technology Transfer Experience at Scanditronix,} Proceedings of the fourth
  European Particle Accelerator Conference (EPAC) in London (1994) 350; online
  available from \url{https://accelconf.web.cern.ch/e94/PDF/EPAC1994_0350.PDF}.}.
Over the years Scanditronix split into several specialized companies that each
focus on those activities. One of them still manufactures smaller cyclotrons for
medical applications in Uppsala today. 
\subsection*{MAXLAB in Lund}
%
Around 1953, KTH in Stockholm delivered a small synchrotron to the
nuclear physics department at Lund university. Colloquially, this machine was referred
to as ``Ur-MAX.'' It accelerated electrons to energies of up to 35\,MeV and crashed them
into a target to produce so-called ``bremsstrahlung,'' in other words, highly energetic
photons, also called gamma rays, suitable to explore atomic nuclei. These photons were
then used to excite various nuclei in order to  determine their energy levels. About ten
years later, in 1962, a larger machine, the ``Lund University SYnchrotron''
LUSY\footnote{For a historical overview see {\em Sagan om Ringen,} available from
  \url{http://history.fysik.lu.se/images/FysicumsHistoriaBok_pdf/SV_FysikILund_web/SV_Bok_20_SOR_web.pdf}
  (in Swedish). A more technical
  narrative about the history of Maxlab is provided by N. Martensson, M. Ericsson,
  {\em The saga of MAX IV, the first multi-bend achromat synchrotron light source,}
  Nuclear Instruments and Methods A 907 (2018) 97.} accelerated electrons to
1200\,MeV in a 34\,m long ring. It served much the same purpose as its smaller predecessor,
namely to explore the physics of atomic nuclei. LUSY operated until 1972 when funding dried
up as a consequence of increased Swedish contributions to international projects.
\par
In order to maintain some local experimental facilities, Lund University contemplated
the construction of a smaller accelerator that would satisfy both the nuclear physics
users but also the emerging synchrotron radiation community that explored the structure
of atoms and molecules. In order to reach higher energies, electrons were first accelerated
to 100\,MeV in a small recirculating linear accelerator, a so-called racetrack microtron. 
These electrons were then passed on to a ring with a circumference of 32\,m that accelerated
the electrons further, all the way up to 550\,MeV. By 1985 this machine, later dubbed MAX~I,
was up and running and indeed served a double purpose. For the nuclear physics users it
provided a stream of electrons that could be used directly or be converted to gamma rays
in a target. And for the synchrotron users lower-energy photons in the soft-X-ray range were
generated in the deflecting magnets and made available through windows in the beam pipe.
\par
Soon after MAX~I was operational a larger ring, entirely dedicated to the production
of synchrotron light, was discussed because the demand for more photons among synchrotron
radiation users was growing. Especially higher photon energies were requested and
generating those required higher electron beam energies. The end point of this discussion
was the construction of MAX~II with a circumference of 90\,m. By 1995 it reached electron
energies of 1500\,MeV. Placing special magnets, so-called undulators and wigglers, in the
ring significantly enhanced the quantity and also the quality of the emitted radiation.
Even X-rays could be served to the user community that was now able to analyze molecules
important for biological and medical applications. But the number of places to install
undulators and wigglers was rather limited and demand for photons, both from Scandinavia,
but increasingly also from international users was still growing. Therefore a third ring,
MAX~III, was planned and eventually constructed.
\par
MAX~III served a double purpose. Operating at an electron energy of 700\,MeV, it
would provide lower-energy photons to those users that required them. At the same time
it would serve as a test bed to develop the technology for a future, and much larger,
accelerator. In order to vastly improve the beam quality of the electrons, Mikael
Ericsson, the head of the accelerator group in Lund, chose an ingenious way of arranging
the magnets. But that required to place large numbers of components in heavily constrained
spaces. To make this possible, he invented a completely new way of manufacturing the
system by carving parts of the magnets from a solid piece of iron with the help of
computer-controlled milling machines. This dramatically reduced the manufacturing costs
and, at the same time, increased the precision of the parts, such that MAX~III could
be constructed in a very economical way. By 2005 MAX~III was ready to operate and started
delivering photons to the users.
\par
Immediately after MAX~III was operating, the accelerator builders in Lund started to
work on a proposal to build a much larger facility that was to be called MAX~IV. The
premises that housed the earlier accelerators was too small, such that a new site
outside of Lund was chosen to become the home of MAX IV. The facility sports a
300\,m long linear accelerator to increase the energy to 3000\,MeV before the electrons
are injected into the larger of two rings. It has a circumference of 528\,m and stores electrons with an
energy of 3000\,MeV. A large number of wigglers and undulators produce synchrotron
light with a large variety of energies that are directed to experimental stations,
where it irradiates samples. In some cases the radiation
knocks out electrons whose energy reveals details about the atomic structure of the
samples. The new MAX~IV facility comprises a second, smaller ring. It has a
circumference of 96\,m and stores electrons with an energy of 1500\,MeV. Like in
the bigger ring, undulators and wigglers generate radiation that
is directed to experimental stations. At the end of the linear accelerator a
dedicated undulator produces very short flashes of radiation that are only possible
with a linear accelerator. The MAX~IV facility started to operate in 2017.
\par
Several of the sixteen experimental stations are devoted to determine the structure
of proteins and pharmaceuticals. Others focus on understanding and optimizing
catalytic materials that help to make certain chemical reactions more efficient---the
catalytic converters installed in most cars may serve as an example. Still others optimize
the performance of solar cells and batteries. For example, finding ways to replace lithium
by the much more abundant sodium is high on the agenda. Many of the relevant chemical
processes depend on the chemistry close to interfaces or surfaces of materials, which
is the prime object of investigation for some of the stations. Rather than covering
the full scientific output, which comprises more than 200 scientific publications in
2023 alone\footnote{We refer to the annual reports available from
  \url{https://www.maxiv.lu.se/science/reports/} to obtain an impression about the
  scientific output of MAX-IV.}, we move a few hundred meters and have a look at
MAX-IV's neighbor, the European Spallation source.
\subsection*{European Spallation Neutron Source  in Lund}
Already in the late 1970s Germany contemplated the construction of a large neutron
source, but the political climate at the time was not favourable and the project was
abandoned. It resurfaced in the late 1990s in the UK, then moved back to Germany
and on to France in the early 2000s. Only towards the end of the decade a pan-European
attempt was successful; three countries sought to host the facility. Finally, in
2009 the European research ministers reached a consensus to build the ``European
Spallation Source'' ESS in Lund. Once finished, it will be the most powerful source
of neutrons to explore new materials as well as new biologically and medically
important substances. In contrast to synchrotron radiation, which mainly interacts with
the electrons in matter, neutrons interact with the atomic nuclei and thus provide
a complementary view of matter on atomic scales. 
\par
The ESS will accelerate a huge number of protons, the nuclei of hydrogen atoms,
in a 500\,m long superconducting linear accelerator to an energy of 2000\,MeV. At
this point the power of all the accelerated protons is huge. 5\,MW of beam crash
into a rotating tungsten wheel where the protons create neutrons in so-called
spallation reactions. This wheel is embedded in a big block of material, called
a moderator, that slows down the neutrons. It might sound strange that we first
accelerate the protons and then have to slow down the neutrons.
However, a high proton energy is needed to produce a sufficient number
of neutrons. On the other hand, only slow neutrons have the ability to resolve the
structure of matter on the atomic scale, because their quantum-mechanical wavelenth
has the same magnitude as the spacing of atoms in matter. In order to extract the
neutrons, the moderator has a number of holes where the slowed-down neutrons can
escape towards fifteen instruments.
\par
The distance from the the target to the instruments is usually on the order of
100\,m or more and that makes it possible to select the speed of the neutrons---and
thus their wavelength---by their arrival time; just briefly open a gate to let the
desired neutrons pass. In some instruments the neutrons are used to make images of
the inside of specimen, in much the same way X-rays provide images of the inside of
our body. This type of analysis is of particular relevance to the manufacturing industry
because it helps to identify material faults and fatigue due to internal stresses and
deformation of samples. In other instruments, the neutrons are scattered from the
surface of samples at a shallow angle. The distribution of reflected neutrons
provides information about ``things'' adsorbed to the surface. Even the structure
of large biological molecules such as proteins can be determined with neutrons.
\par
Though the production of copious numbers of neutrons is the prime task of the ESS,
ideas emerged to use the highly intense proton beam to
produce neutrinos. But this requires a substantial upgrade of the facility to
produce neutrinos concurrently with neutrons; the linac has to accelerate twice
as much beam. Moreover, a moderately large ring has to wind up long beam pulses
and eject them as a much shorter, but also more intense pulse onto a second target
where a sequence of nuclear processes converts the impinging protons to neutrinos.
And finally, large detectors for the neutrinos have to be constructed, usually deep
below ground level in underground mines.
\subsection*{And there is more}
After the the war, in the late 1940s, the Swedish ``Atomkommitte'' funded several
electrostatic van de Graaff accelerators operating with MeV energies in Gothenburg,
Lund, Uppsala, and Stockholm. The tandem accelerator, mentioned in the section about
Uppsala, is a descendant of these machines, as is a comparable one in Lund.
\par
Many hospitals operate small electron accelerators for irradiating tumors in patients,
either directly with the electrons, or with X-ray photons. Those are created by first
impinging the electrons onto a target from where the X-rays emerge. Historically,
various types of accelerators such as betatrons, microtrons, and linear accelerators
were used. Today, practically all systems are based on linear accelerators. The university
hospital in Uppsala, for example, operates several of these machines\footnote{Many
  of these electron accelerators are manufactured by the company Elekta in
  Stockholm, a late descendent of the activities at KTH mentioned above.} delivering
electrons with energies of typically 20\,MeV.
\par
Even small cyclotrons\footnote{Such cyclotrons are manufactured by the company
  General Electric Healthcare in Uppsala that has evolved from one of the branches of
  Scanditronix mentioned above.} accelerating protons to about 20\,MeV are installed
in the basement of some hospitals. They create the radioactive isotopes needed to
follow the path of marked molecules through the human body. Some of the isotopes
emit positrons that cause the emission of characteristic X-rays, which allows the
medical doctors to pinpoint their origin in a positron-emission-tomography (PET)
scanner.
\subsection*{Acknowledgements}
I am grateful to my colleagues Micke Pettersson and Kjell Fransson for pointers to
old publications in not-so-well-known journals and to other reports. Anders Montelius
provided valuable information about accelerators in hospitals. I gratefully acknowledge
advice and helpful suggestions from Mikael Eriksson, Roger Ruber, and Jan-Erik Rubensson.
\end{document}